\documentclass{aa}
\usepackage[varg]{txfonts}
\usepackage{amsmath}
\usepackage{amsfonts}
\usepackage{graphicx,xcolor, soul}
\usepackage{subcaption}
\usepackage{url}
\usepackage[normalem]{ulem}

\bibpunct{(}{)}{;}{a}{}{,} 

\usepackage[utf8]{inputenc}
\usepackage{threeparttable}
\usepackage{hyperref}   
\hypersetup{colorlinks=true,linkcolor=blue,citecolor=blue,filecolor=blue,urlcolor=blue}
\usepackage{cleveref}
\usepackage{threeparttable}
\usepackage{comment} 
\usepackage{xcolor}
\def\la{\raise.5ex\hbox{$<$}\kern-.8em\lower 1mm\hbox{$\sim$}}
\def\ga{\raise.5ex\hbox{$>$}\kern-.8em\lower 1mm\hbox{$\sim$}}
\def\be{\begin{equation}}
\def\ee{\end{equation}}
\def\ba{\begin{eqnarray}}
\def\ea{\end{eqnarray}}
\def\bt{\begin{tabular}{@{}c@{}}}
\def\et{\end{tabular}}

\def\rlc{r_{\mathrm{LC}}}
\def\tpc{\theta_{\mathrm{pc}}}

\def\chin{\chi^2_{\nu}}
\newcommand{\rlight}{r_{\rm L}}

\begin{document}

\title{Constraining millisecond pulsar geometry using time-aligned radio and gamma-ray pulse profile}
\author{Onur Benli$^1$\thanks{E-mail:obenli@unistra.fr} , J\'er\^ome P\'etri$^1$, Dipanjan Mitra$^{2,3}$
}

\institute{$^1$ Universit\'e de Strasbourg, CNRS, Observatoire astronomique de Strasbourg, UMR 7550, F-67000 Strasbourg, France \\
$^{2}$ National Centre for Radio Astrophysics, Tata Institute for Fundamental Research, Post Bag 3, Ganeshkhind, Pune 411007, India\\
$^3$ Janusz Gil Institute of Astronomy, University of Zielona G\'ora, ul. Szafrana 2, 65-516 Zielona G\'ora, Poland
}

\titlerunning{Constraining MSP geometry}
\authorrunning{Benli, P{\'e}tri, Mitra}
\date{2020}

\def\refbf{}

\abstract{

{\textit Context.} Since the launch of the Fermi Gamma-Ray Space Telescope, several hundred gamma-ray pulsars have been discovered, some being radio-loud and some radio-quiet with time-aligned radio and gamma-ray light curves. In the second Fermi Pulsar Catalogue, 117 new gamma-ray pulsars have been reported based on three years of data collected by the Large Area Telescope on the Fermi satellite, providing a wealth of information such as the peak separation~$\Delta$ of the gamma-ray pulsations and the radio lag~$\delta$ between the gamma-ray and radio pulses.

{\textit Aim.} We selected several radio-loud millisecond gamma-ray pulsars with period~$P$ in the range 2-6~ms and showing a double peak in their gamma-ray profiles. We attempted to constrain the geometry of their magnetosphere, namely the magnetic axis and line-of-sight inclination angles for each of these systems.

{\textit Method.} We applied a force-free dipole magnetosphere from the stellar surface up to the striped wind region ---well outside the light cylinder--- to fit the observed pulse profiles in gamma-rays, consistently with their phase alignment with the radio profile. In deciding whether a fitted curve is reasonable or not, we employed a least-square method to compare the observed gamma-ray intensity with that found from our model, emphasising the amplitude of the gamma-ray peaks, their separation, and the phase lag between radio and gamma-ray peaks.

{\textit Results.} We obtained the best fits and reasonable parameters in agreement with observations for ten millisecond pulsars. Eventually, we constrained the geometry of each pulsar described by the magnetic inclination~$\alpha$ and the light-of-sight inclination~$\zeta$. We found that both angles are larger than approximately~$45^{\rm o}$.
}


\maketitle

\section{Introduction}

The launch of the Fermi Gamma-Ray Space Telescope in June~2008 revolutionised our knowledge of the physics of pulsars when it discovered hundreds of new gamma-ray pulsars, among them many millisecond ones \citep{abdo_second_2013}, a surprising result at that time. The second gamma-ray pulsar catalogue (2PC) contains 117~pulsars. The most recent catalogue, 3PC, will soon be reported, now containing 300~pulsars. These gamma-ray pulsars are either radio-loud or radio-quiet. If seen in both wavelengths, their radio and gamma-ray light curves provide a wealth of information about their magnetospheric topology and emission sites close to the neutron star surface and in regions near the light-cylinder radius ($\rlc$), respectively. The particles are likely to corotate with the neutron star within $\rlc$ and flow beyond $\rlc$ as a pulsar wind, that is, a relativistic magnetised outflow of pair plasmas, forming a current sheet known as the striped wind \citep{coroniti_magnetically_1990, michel_magnetic_1994}.

Particle trajectories and photon-production mechanisms within the magnetosphere and the electromagnetic field topology itself are all intertwined with each other. Realistic modelling of these quantities therefore necessitates the computation of the plasma-radiation field interaction in a self-consistent manner. Several promising approaches in this direction have been undertaken by many authors using either a fluid description ---see for example \cite{spitkovsky_time-dependent_2006}, \cite{bai_modeling_2010}, \cite{kalapotharakos_gamma-ray_2012}, \cite{petri_unified_2011, petri_high-energy_2012, petri_general-relativistic_2018}--- or a particle description; see for example \citep{cerutti_modelling_2016}. The vacuum-retarded dipole solution \citep{deutsch_electromagnetic_1955} entails vanishing particle density (no particle to accelerate) while possessing a huge electric field component parallel to the magnetic field inside the magnetosphere. Force-free electrodynamics (FFE) solutions, on the contrary, must have large values of the charge density, which shorts out the parallel component of the electric field and precludes the particle acceleration. Some resistivity or dissipation is required to avoid a full screening of the electric field, allowing for efficient particle acceleration to ultra-relativistic speeds.

Another complication while modelling pulsar electrodynamics stems from the magnetic obliquity with respect to the spin axis of the pulsar. Initial attempts to understand the physics underlying the pulsar magnetosphere commonly focused on the aligned rotator assumption. The reason was that finding a solution for the aligned case is much easier than for the more realistic oblique case. \cite{contopoulos_axisymmetric_1999} produced successful numerical simulations to encounter FFE for the first time for the aligned rotator. Several authors constructed a force-free pulsar magnetosphere by considering time-dependent simulations for the oblique case \citep{spitkovsky_time-dependent_2006, kalapotharakos_three-dimensional_2009, petri_pulsar_2012}. Because the simulations would otherwise be very time consuming, in these previous works, a large ratio of $R/\rlc$ = 0.2-0.3 was used corresponding to an unrealistically fast pulsar with periods of $\sim 1$~ms. Some authors proposed models including an `electrosphere', in which a differentially rotating plasma flows around the equatorial plane, separated from the oppositely charged domes by huge gaps  \citep{krause-polstorff_electrosphere_1985, petri_global_2002, mcdonald_investigations_2009}.
  
For the first time, \cite{watters_atlas_2009} performed a detailed study of pulsar light curves with an individual assessment of polar cap, outer gap, and slot gap models. Since then, several authors have contributed to the goal of producing a complete atlas of pulsar light curves for different obliquities~$\alpha$ and inclination angles~$\zeta$. \cite{venter_probing_2009}, \cite{johnson_constraints_2014}, and \cite{pierbattista_light-curve_2015} adopted models based on the vacuum-retarded dipole solution, first given by \cite{deutsch_electromagnetic_1955}, and later extended to include rotational sweep-back magnetic field lines (e.g. \citealt{dyks_relativistic_2004}). \cite{contopoulos_pulsar_2010} studied the emission of curvature radiation from the current sheet that develops around the equatorial plane in the force-free regime, which they called pulsar synchrotron, and were able to reproduce the high-energy light curves emitted by pulsars.

Estimation of the radio and gamma-ray time lag~$\delta$ and gamma-ray peak separation~$\Delta$ for those pulsars showing double-peaked light curves and application to the observed light curves of several gamma-ray pulsars provided in the first Fermi Catalogue \cite{abdo_first_2010} was carried out by \cite{petri_unified_2011}  who used an oblique split monopole solution \citep{bogovalov_physics_1999} for the field structure and a simple polar cap geometry for the radio counterpart. With the study presented here, our aim is to understand the characteristics of the gamma-ray and radio light curves by employing the FFE prescription where we also take into account the real effect of particle flow within the dipolar magnetosphere on the field structure. \cite{contopoulos_pulsar_2010} also compared the predicted relation $\delta-\Delta$, assuming emission near the current sheet, to the calculated relations based on the first pulsar catalogue. \cite{kalapotharakos_gamma-ray_2012} focused on the orthogonal rotator case, taking into account a large range of conductivity in dissipative pulsar magnetospheres. In \cite{kalapotharakos_gamma-ray_2014}, the orthogonal condition was relaxed to the whole range of magnetic obliquity to produce $\delta$ and $\Delta$ measured by the observations. By comparing the observed $\Delta - \delta$ diagram with their estimations from the model, these authors favoured the cases with high conductivities; as the conductivity increases, the system converges to the FFE limit.    

\cite{chen_numerical_2020} recently suggested a force-free magnetospheric model and presented radio, X-ray, and gamma-ray light curves in agreement with the observations of the millisecond pulsar (MSP) PSR~J0030+0451, providing an estimation of the magnetic inclination of the pulsar of $\sim 80^{\circ}$. More recently, \cite{kalapotharakos_multipolar_2020} proposed the FFE model by taking into account an off-centred dipole field and quadrupole components, the Kerr-metric for the ray-tracing of photons from a distant observer to the hot spots on the surface of the star. 

Fitting the radio-loud gamma-ray pulsar high-energy light curve heavily relies on a good time alignment constraint between radio and gamma-rays. Radio profiles are often fitted by a single or multiple Gaussian components. In our approach, outlined below, the radio beam emitted from the polar cap is symmetric, and the intensity profile is defined by a Gaussian centred at the magnetic axis; the fiducial phase for the radio light curve of a pulsar corresponds to the phase at which the magnetic axis lies in the plane defined by the rotation axis and observer's line of sight. For normal (young) pulsars with periods above about 100 ms milliseconds,  the rotating vector model \citep{radhakrishnan_magnetic_1969} is usually exploited to constrain the fiducial phase and emission height. However, this method is inappropriate for MSPs with periods of about a few milliseconds because of the complex structure of their pulse shapes. This is mainly due to the fact that radio emission emanates from altitudes of a substantial fraction of the light-cylinder radius where relativistic effects are important (aberration, magnetic sweep-back, plasma currents), and very close to the stellar surface where multipolar magnetic field components are expected to be important. Therefore, the phase-zero determination of the radio light curve for MSPs is problematic, not to mention that the radio duty cycle can also be very large. Therefore, for the definition of the phase zero of ten pulsars investigated in this work, two distinct methods were used in the second Fermi Pulsar Catalogue: one taking into account the fiducial point at peak intensity; and the other using the fiducial point placed at the point of symmetry of the prominent radio peak. PSR~J0030+0451, PSR~J0102+4839, PSR~J0437--4715, PSR~J1614--2230, PSR~J2017+0603, and PSR~J2043+1711 belong to the former category while PSR~J0614–3329, PSR~J1124–3653, PSR~J1514–4946, and PSR~J2302+4442 belong to the latter category (see Table 8 in \citealt{abdo_second_2013}). In this study, we adopted the same fiducial phases as those given in the catalogue. Independently of the method used in the catalogue, all zero phases correspond to the alignment of the magnetic axis with the line of sight in our model.

All these recent studies are very promising for the constraint of the geometry of the pulsar and are the subject of this paper. The paper is organised as follows. In Section~\ref{sec:Radio} we briefly mention the radio emission phenomenology of pulsars and draw attention to the complications in the radio pulse profiles of MSPs. In Section~\ref{sec:Gamma}, we summarise the gamma-ray observations of Fermi/LAT presented in 2PC. In Section~\ref{sec:FFE}, we describe the force-free model for the magnetosphere and the striped wind where gamma-ray pulses are assumed to be produced in our model. We explain our fitting method in Section~\ref{sec:Fitting}. Remaining consistent with radio and gamma-ray time-alignment, we reproduced the observed gamma-ray pulse profiles for the sample of MSPs and attempted to determine the geometry of each pulsar. We present our results in Section~\ref{sec:Results}. A broader discussion of our method is provided in Section~\ref{sec:Discussion}. We draw conclusions in Section~\ref{sec:Conclusion}.

\section{Radio observations and phenomenology}
\label{sec:Radio}

We aim to use radio and gamma-ray observations to find the dipolar magnetic field emission geometry of the pulsar with respect to the observer. The role of the radio observations enters the problem in the following manner. Firstly, one needs to assume that the radio emission arises deeper in the magnetosphere closer to the surface of the neutron star and secondly the emission needs to arise from regions where the magnetic field geometry can be closely approximated with a static dipole.
These two assumptions can then be used to treat the radio emission as a proxy for identifying the fiducial dipolar  plane of the magnetic field lines; the geometry of the gamma-ray emission can then be solved with respect to this plane. The assumptions hold true for normal-period pulses (i.e. pulsars with periods longer than about 100~ms), where detailed phenomenological pulse profile and polarisation studies suggest that the radio pulsar emission beam comprises a central core emission surrounded by two
nested conal emissions (see \cite{rankin_toward_1983, rankin_toward_1990, rankin_toward_1993, mitra_revisiting_1999}.  The emission arises from  regions of open dipolar magnetic field lines below 10\% of the light-cylinder radius; see \cite{mitra_comparing_2004} and \cite{mitra_nature_2017}. These conclusions were based on demonstrations that the observed opening angle of normal pulsars follows the $P^{-0.5}$ (where $P$ is the pulsar period) behaviour (see, e.g. \citealt{skrzypczak_meterwavelength_2018}) which is as expected from emission arising from the open dipolar magnetic field line region. Secondly, the polarisation position angle (PPA) across the pulse profile follows the rotating vector model (RVM), which is an indication of emission arising from regions of diverging magnetic field line geometry. Further, the effect of aberration and retardation in the linear approximation has been observed in normal radio pulsars ---see \cite{blaskiewicz_relativistic_1991} \cite{petri_joint_2020}---, whereby a shift between the centre of the total intensity profile and the steepest gradient point of the position angle curve is observed. This shift is proportional to $r_{\rm em}/\rlight$ and is used to obtain the radio emission heights.

However, in MSPs, the normal period pulsar phenomenology cannot be applied. A radio pulse profile does not have any orderly behaviour, and the PPA  traverse is significantly more complex and cannot be modelled using the RVM. Only recently, \cite{rankin_toward_2017} tried to model a few MSPs in terms of the core cone model and suggested that the radio emission might arise from a few tens of kilometres above the neutron star surface. However, these estimates are highly uncertain, because in none of the cases can the RVM be fitted to the PPA traverse. Therefore, in observations of MSPs, the location of the radio emission region is not constrained. In this study, we acknowledge this drawback and assume that the emission arises from regions anchored on open dipolar magnetic field lines of about 10\% of light-cylinder radius~$\rlight$.

\section{Gamma-ray observations}
\label{sec:Gamma}

The Large Area Telescope (LAT) instrument onboard the Fermi Satellite, launched in June~2008, detects gamma-rays with energies between 20~MeV and $\sim300$~GeV. The full three years of data used to compile the 2PC in \cite{abdo_second_2013} was collected via the observations from 2008 August~4 to 2011 August~4. The gamma-ray pulsar search using the known rotation ephemerides of radio or X-ray pulsars led to the discovery of 61 gamma-ray pulsars in the 2PC. Phase alignment of gamma-ray pulses with radio pulses provides vital information about the geometry of the different emission regions. Another method used in the 2PC for discovering gamma-ray pulsars is the blind periodicity search which can provide ephemeris of pulsars relying only on the LAT gamma-ray data. In the catalogue, 36 new gamma-ray pulsars were discovered with this method. In addition, radio searches of several hundred LAT sources by the Fermi Pulsar Search Consortium provided 47 new pulsar discoveries, including 43 MSPs and 4 young or middle-aged pulsars \citep{ray+_2012}. One of the most prominent differences of the 2PC from the 1PC is the much higher ratio of MSPs to the normal pulsar population.

Spectra of gamma-ray pulsars are relatively insensitive to the neutron star period, peaking around a cut-off energy of several GeV for all of them. Moreover, the subexponential decrease of the spectra above this cut-off rules out the polar cap scenario for gamma-ray emission because of the strong magnetic opacity at these energies \citep{abdo_fermi_2009-1}. Slot gaps and outer gaps have since been the favoured site of high-energy photon production. While there is no independent observational constraint on the location of the gamma-ray emission, because the radio time lag~$\delta$ and peak separation~$\Delta$ values found in 2PC are supported by extensive force-free and dissipative magnetosphere simulations, these regions are shifted more and more towards the light cylinder and even outside in the striped wind \citep{petri_unified_2011}. Therefore, in the present study, we use a magnetosphere model (see section~\ref{sec:FFE}) where the gamma-ray emission starts from the light cylinder and continues into the striped wind.

\section{Magnetosphere and emission model}
\label{sec:FFE}

As the neutron star is surrounded by plasma made of electron--positron pairs producing the observed radiation, we need an accurate solution for the neutron star magnetosphere. Therefore, the radio and gamma-ray light-curve computations rely on the electromagnetic field structure extracted from the force-free pulsar magnetosphere obtained by time-dependent numerical simulations. Such calculations have been performed by \cite{petri_electrodynamics_2020} using his time-dependent pseudo-spectral code detailed in \cite{petri_pulsar_2012}. Specifically, the neutron star radius is set to $R/\rlight=0.2$ corresponding to a 1.2~ms period and the artificial outer boundary is set at $r=7\,\rlight$ where the light-cylinder radius is $\rlight=c/\Omega$ and $\Omega=2\,\pi/P$. The numerical solution of the electromagnetic field is therefore expressed as a series in Fourier-Chebyshev polynomials and can be accurately evaluated at any arbitrary point within the simulation box. The striped wind starting from the light cylinder is also self-consistently computed on almost one wavelength $\lambda_{\rm L} = 2\,\pi\,\rlight$.

\begin{figure*} 
        \centering
        \includegraphics[width=2.\columnwidth,angle=0]{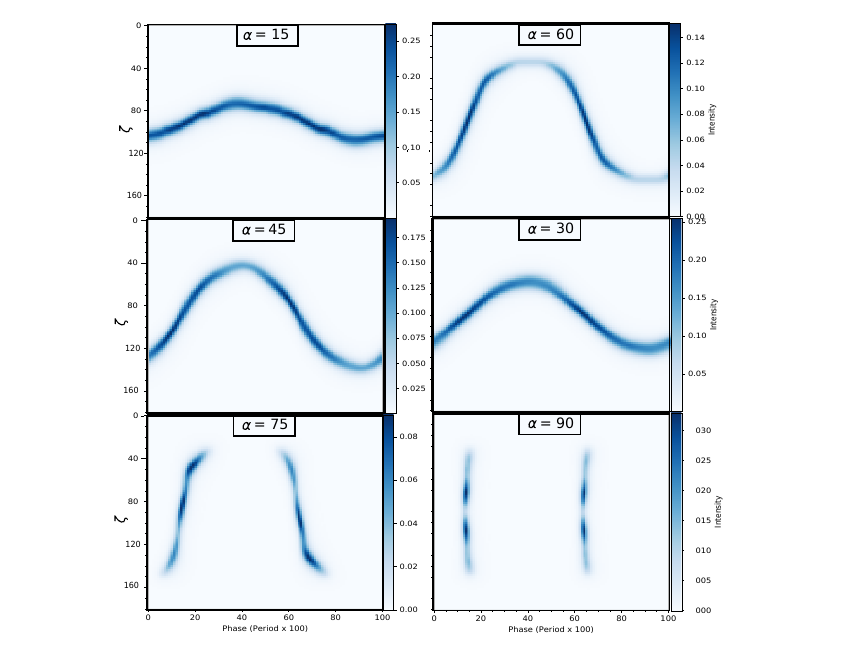}
        \caption{Sky-maps for $\alpha = 15^{\circ}, 30^{\circ}, 45^{\circ}, 60^{\circ}, 75^{\circ}$ and $90^{\circ}$. The x-y axes denote the rotation phase (in the unit of $100 \times P$) and $\zeta$ for each panel while the intensity of gamma-rays (in arbitrary units) is indicated by the colour bar on the right of each panel. 
        }
        \label{fig:lightcurves}
\end{figure*}

The two emission sites, radio and high-energy, are described as follows. Radio photons are expected to emanate from the open field-line regions above the polar caps whereas the gamma-ray photons are produced in the current sheet of the striped wind, outside the light cylinder at $r\geq\rlight$. The emissivity of the striped wind drops sharply with distance, so we only considered emission in the shell $r\in[1,2]\,\rlc$. We extended our previous analysis made by \cite{petri_unified_2011} where a split monopole wind was assumed to be directly connected to the stellar surface. Now this approximation, which in reality does not hold inside the light cylinder, is replaced by a realistic force-free dipole magnetosphere smoothly linking the quasi-static zone inside the light cylinder to the wind zone outside the light cylinder. Figure \ref{fig:lightcurves}  shows the sky-maps produced by our striped wind model for six different magnetic inclination angles~$\alpha$. In each panel, the phase of prominent radio peak is set to zero, and gamma-ray light curves are plotted according to a time-alignment shift of the phase for synchronisation with the radio pulse profile. 

The gamma-ray pulse width is a combination between the relativistic beaming effect and the depth of the line of sight crossing the current sheet. For infinitely thin current sheet, this width~$W_\gamma$ scales as the inverse of the wind Lorentz factor~$\Gamma$ and $W_\gamma \propto 1/\Gamma$ as shown by \cite{kirk_pulsed_2002}. If the thickness is finite, it will reach a lower limit reflecting the current sheet thickness scaled to the wind wavelength $\lambda_{\rm L}$ for ultra-relativistic speeds~$\Gamma\gg1$. This has been observed in the Geminga pulsar as reported by \cite{abdo_fermi-lat_2010}. The gamma-ray peak intensity for each pulse also depends on the length of the line of sight crossing the current sheet in each pulse. Contrary to the split monopole case, the striped wind geometry is less symmetric because of the distorted dipole due to rotation and plasma effects. This loss of symmetry imprints the associated pulsed profile, which is also less symmetric, showing discrepancies in the peak intensities compared to the split monopole emission.

The sky maps are constructed as follows. First, we determine the polar cap rims by identifying the last closed field lines grazing the light cylinder. These field lines also draw the base of the current sheet in the striped wind. More precisely, radio emission is assumed to occur over the whole polar cap area at $r=R$ (therefore at zero altitudes in our simulation box). As we are 
not interested in the exact radio pulse profiles, which often obtain a complex shape with multiple components, but only in the location of the maximum intensity taken as phase zero for time-alignment purposes do we modulate the polar cap emission with a Gaussian profile~$w$ peaking at the magnetic axis such that
\begin{equation}\label{eq:poids}
 w = e^{4\,((\mu\cdot\mathbf{n})^2-1)}
,\end{equation}
where $\mu$ is the magnetic moment vector and $\mathbf{n}$ the unit vector pointing in the direction of the line of sight. This profile is not intended to reproduce exact pulse profiles but simply to easily locate the maximum of the radio peak that sets the phase zero for the radio and gamma-ray light curves. Emission in the striped wind is located around the current which is identified by the location outside the light cylinder where the radial magnetic field component reverses sign. In an ideal picture, this current sheet is infinitely thin, but due to dissipation, it will spread to a certain thickness depending on the plasma condition. The variation of the Gaussian profile does not affect the phase of prominent radio or gamma-ray peaks in our model. It only impacts on the width of the radio pulses. However, an exact fitting of the radio pulse profiles is out of the scope of this work; we only use its peak location to determine the phase zero for time alignment purposes. Moreover, by construction in our model, the radio peak intensity corresponds to the phase of the magnetic axis (north pole or south pole). We stress that only the radio pulse phase is needed, regardless of its shape, to constrain gamma-ray peak phases and thus the geometry of pulsars.

The dipolar polar cap size has a half opening angle of $\theta_{\mathrm{pc}} \approx \sin \theta_{\mathrm{pc}} = \sqrt{R/\rlc}$. This corresponds to a radio pulse profile width of $W \approx 3/2\,\theta_{\mathrm{pc}}$ ,  which also takes into account the magnetic field line curvature, assuming a static dipole which is a good approximation at the surface as long as $R\ll\rlc$. For our 1.2~ms pulsar computed from the simulations, this corresponds to a width of $W \approx 38^{\circ}$, but for the sample of MSPs we have fitted below, with a 3~ms period  on average, this width shrinks to $W \approx 25^{\circ}$. 
Therefore, when the observer line of sight~$\zeta$ lies within the radio emission beam ($\alpha - W \leq \zeta \leq \alpha + W$),  a radio pulse will still be detected. We added this constraint in the fitting procedure explained in the following section.

\section{Fitting method}
\label{sec:Fitting}

The radio and gamma-ray light curves of the pulsars are produced by numerical calculations assuming a force-free magnetosphere. To produce a complete sky-map, we trace the inclination angle between the magnetic dipole moment and the rotation axis from $\alpha = 0^{\circ}$ to $90^{\circ}$ with $5^{\circ}$ resolution, for the inclination of the line of sight with respect to the rotational axis ($\zeta$) varying from $0^{\circ}$ to $180^{\circ}$ with 2$^{\circ}$ resolution. The maximum intensities for radio and gamma-rays are independently normalised to unity. We do not model the particle distribution within the magnetosphere with a self-consistent physical approach. In that sense, our model is purely geometric. The gamma-ray intensity is a free parameter of the model while the form of the light curves is obtained by solving the equations numerically for the force-free electrodynamics (see Section \ref{sec:FFE}). We are mainly interested in the time lag~$\delta$ between the radio and the first gamma-ray peak, the gamma-ray peak separation $\Delta$ and the ratio of the peak amplitude. These are geometric constraints, and hence we cannot model the exact pulse profile shapes.

In our analyses, we take into account the radio pulse profiles provided in the 2PC to reproduce gamma-ray pulse profiles calculated by our model with the consistent time lags between gamma-ray and radio peaks. The radio observations of PSR~J0030+0451, PSR~J0614--3329, PSR~J1614--2230, PSR~J2017+0603 and PSR~J2302+4442 by the Nançay Radio Observatory (1.4~GHz band), of PSR~J0102+4839 (1.5~GHz band) and PSR~J1124--3653 (0.8~GHz band) by the Green Bank Telescope, of PSR~J0437--4715 and PSR~J1514 by the Parkes Radio Telescope (1.4~GHz band), and of PSR~J2043+1711 by the Arecibo Observatory (1.4~GHz band) were used in the 2PC; see \citealt{abdo_second_2013} and the ATNF Pulsar Catalogue\footnote{http://www.atnf.csiro.au/research/pulsar/psrcat} (\citealt{manchester_australia_2005}).

\begin{figure} 
\centering
\includegraphics[width=1.1\columnwidth,angle=0]{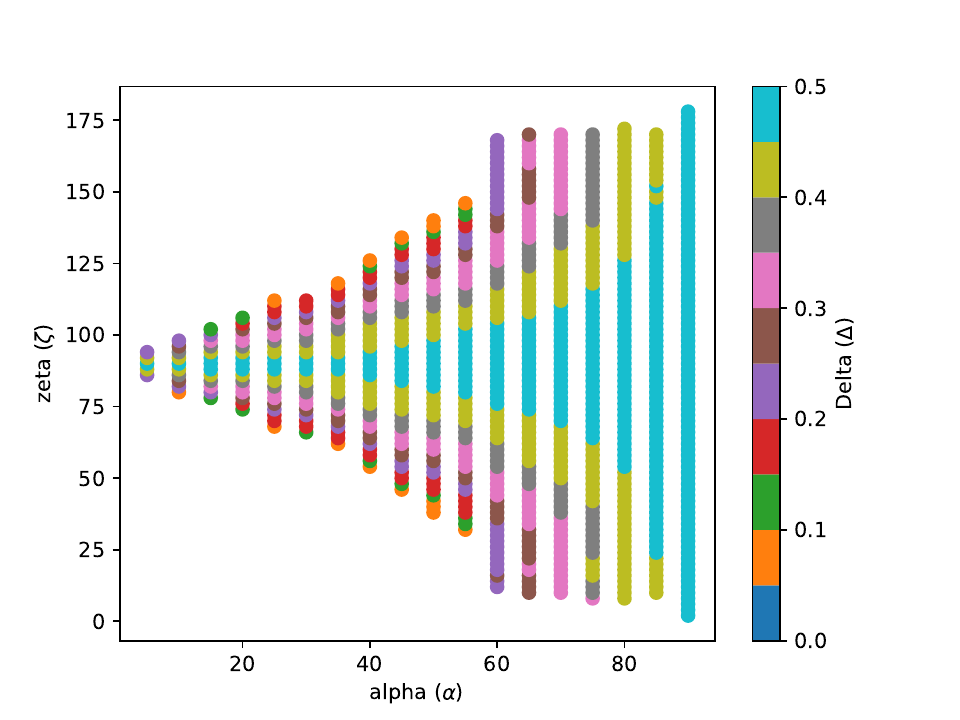}
\caption{Peak separation distribution for the entire space of $\alpha - \zeta$. 
}
\label{fig:Delta}
\end{figure}

We set the time of maximum radio intensity to phase zero and shifted the gamma-ray profile accordingly. In our model, there is a lag~$\delta$ between the gamma and radio peaks up to 0.5 in phase. For $\zeta$ and $\alpha$ values producing double-peaked gamma-ray light-curves in the direction of the observer, we measure the phase difference of these peaks (peak separation, $\Delta$). The expected peak separation measured from our model for a given set $\zeta - \alpha$ can be seen in Fig. \ref{fig:Delta}. The maximum peak separation is 0.5 in units of the pulsar period~$P$. We note that the distribution of the peak separation in the force-free model is compatible with that found from the asymptotic split-monopole solutions \citep{petri_unified_2011}.  

For each source, in order to determine the goodness of fits to the observed gamma-ray light curve, we use a simple least-square procedure, as follows, and try to constrain the reasonable $\alpha$ and $\zeta$ values that produce the best-fitting model curves. The reduced $\chi^2$ is defined by
\be
\label{eq:chi2}
\chi^{2}_\mathrm{\nu}=\frac{1}{\nu}\sum \frac{\left(I^\mathrm{obs}_\mathrm{i}-I^\mathrm{model}_\mathrm{i}\right)^{2}}{\sigma^2_\mathrm{i}} 
,\ee
where $\nu$ is the degree of freedom, which is equals to the number of data points minus the number of fitted parameters, $I^\mathrm{obs}_\mathrm{i}$ and $\sigma^2_\mathrm{i}$ are the observed intensity and associated error of gamma-rays for the $i^\mathrm{th}$ phase bin, and $I^\mathrm{model}_\mathrm{i}$ is the model prediction at the observational phase bin. As the phase space in the model outputs is evenly distributed from 0 to 1 with intervals of 0.01, we interpolate the model light curve and find the intensity at the phase of observation so that we can apply the least-square analysis to the intensities. Regarding the observation data, we take the lower bounds for each bin in the light curve as the time of observation of the associated intensity. As the difference of lower and upper bounds for each bin is sufficiently marginal, e.g. 0.01 for gamma-ray and smaller for the radio light curves, using the minimum phase of each bin instead of the mean phase does not change the fit quality significantly. Also, we subtract the nominal value for the background (reported in the 100~MeV - 100~GeV energy band for each source) from the total intensity in order to remove background noise effects.  


We mainly focus on the sources with clearly measured gamma-ray and radio light curves with high signal-to-noise ratios. Of the numerous sources observed both in gamma-rays and radio bands and reported in the secong Fermi Catalogue \mbox{\citep{abdo_second_2013}}, we pick up pulsars with (1) double-peaks observed in gamma-rays for the complete phase of the pulsar, (2) a well-described prominent peak in the radio profile and clear peak locations, and (3) a gamma-ray peak detection time which is not very close to that of the prominent radio peak in order to obtain their gamma-ray peak phases (localised according to their radio peaks). We should note that the gamma-ray peak properties of pulsars, which do not suit item (3) given above, could not be reproduced by our model. Some freedom in the emission height could add retardation effects, shifting the gamma-ray light curves in the right direction. However, this is at the expense of adding a new free parameter in our model, which we want to avoid in a first attempt to constrain the geometry.

As mentioned earlier, to obtain the geometry for our sample pulsars, we fit the observed gamma-ray light curves with the model light curves using the $\chin$ minimisation process given by equation~\eqref{eq:chi2}. We note that in our model there are three factors that influence the geometry:
(1) the height ratio of the gamma-ray peaks (amplitude), (2) the gamma-ray peak separation, $\Delta$, and (3) the phase lag between radio and associated gamma-ray peaks, $\delta$. However, bridge emission between the gamma-ray peaks, as often seen in so many observed gamma-ray light curves, is not predicted by the model.
Therefore, while finding a best fit and calculating a $\chin$ value, although we take into account off-peak gamma-ray data together with bridge emission between gamma-ray peaks, we aim to retrieve the light curves best suited to the three above conditions, $\Delta$, $\delta.$ and peak amplitude ratio. Therefore, we still regard $\chin$ values for the best fit of some pulsars as adequate, even though they are not close to unity. We also examined a weighted fit for the peaks, such as calculating $\chin$ only around peaks or giving more weight to the data around peaks which do not lead to the estimation of different geometry. We therefore apply the unweighted fitting method for all pulsars here with one exception (see Table~\ref{tab:parameters}).

\section{Millisecond pulsars with time-aligned radio and gamma-ray pulse profiles }
\label{sec:Results}

In the framework of force-free electrodynamics, we aimed to fit the observed gamma-ray light-curves (peak properties) and tried to 
reasonably constrain the values of the angles $\alpha$ and $\zeta$ for MSPs with high signal-to-noise ratios. The best fits of ten pulsars were retrieved by minimising the square of the differences between the gamma-ray intensity observed and obtained from the model~\eqref{eq:chi2}, integrated with all phase bins for each pulsar. We show and discuss our results below for each source, separately.

In our model, we assume radio emission heights to be close to the stellar surface of the simulated magnetosphere and the gamma-ray pulsations to start at $\rlc$ up to $2\,\rlc$ for the sake of simplicity. Radio emission height and the exact region where gamma-ray pulsations are produced are subject to a certain amount of uncertainty, giving us some freedom to shift the gamma-ray light curves with respect to the radio peak, denoted by an additional phase shift $\phi$ accounting for a time-of-flight effect reflecting the variation in emission height of less than approximately~10\% in the leading or trailing direction. For each pulsar, we set the offset $\phi$ to the value that gives the best fit to its gamma-ray light curve and plotted reduced chi-squared statistics for different angles of the line of sight and obliquity by adopting this $\phi$. Although the least $\chin$ can be found with any $\alpha - \zeta$, some angles cannot be excepted as good solutions. Possible solutions are restricted by the geometrical conditions: (1) $\alpha - W \leq \zeta \leq \alpha + W$ (we adopt $W \sim 20^{\circ}$ for all pulsars in this work) to be able to observe radio pulsation and (2) $\zeta \simeq \alpha \geq \pi/4$ to see simultaneous radio and gamma-ray pulsations. We assume $\tpc = 20^{\circ}$ for all MSPs investigated in this paper. Also, the retrieved gamma-ray light curves show symmetry in $\zeta$ around $90^{\circ}$. 


\begin{figure*}[]
  \centering
  
  \begin{subfigure}[b]{0.4\linewidth}
    \includegraphics[width=0.95\linewidth]{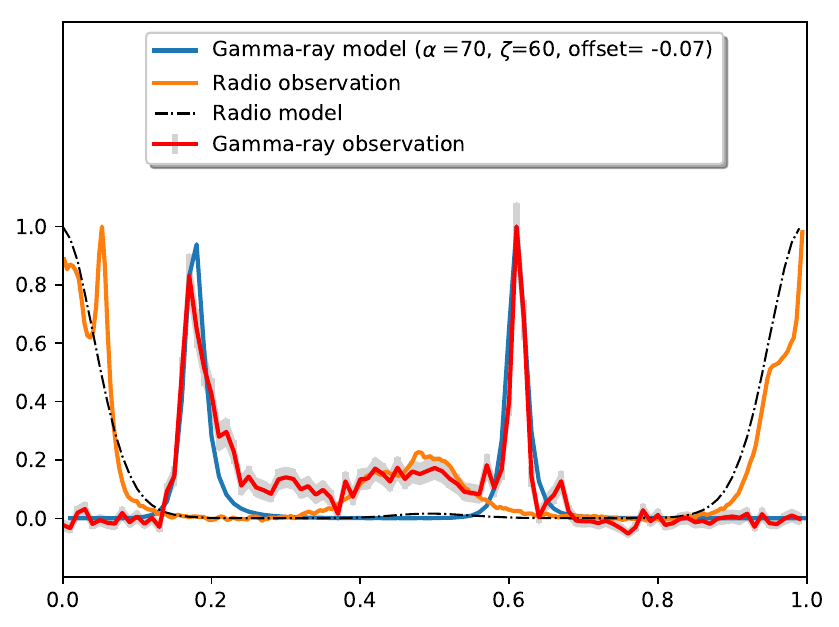}
    \caption{PSR~J0030+0451}
    \label{fig:J0030_bestfit}
  \end{subfigure}
  \begin{subfigure}[b]{0.4\linewidth}
    \includegraphics[width=0.95\linewidth]{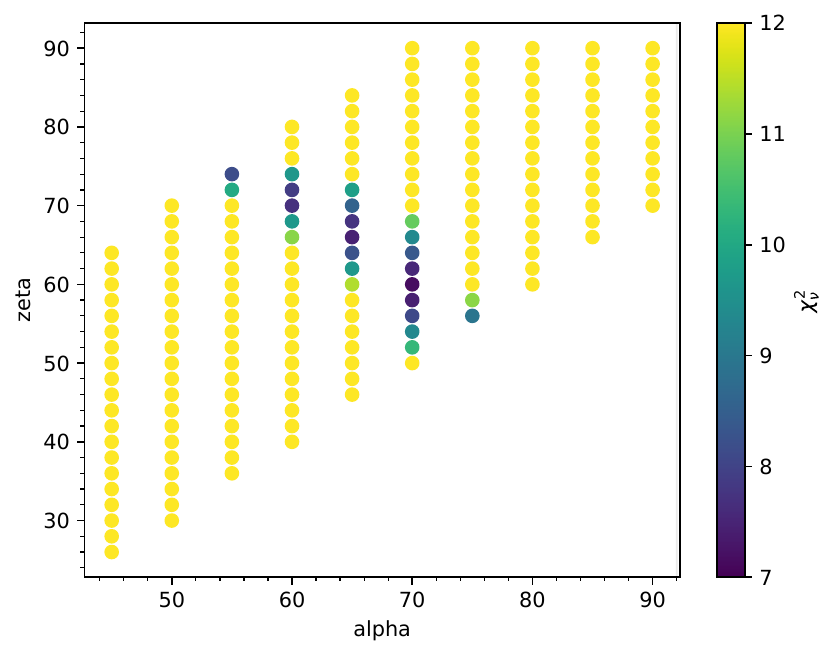}
    \caption{PSR~J0030+0451, $\phi = -0.07$ }
    \label{fig:J0030_chin}
  \end{subfigure}
  
    \begin{subfigure}[b]{0.4\linewidth}
    \includegraphics[width=0.95\linewidth]{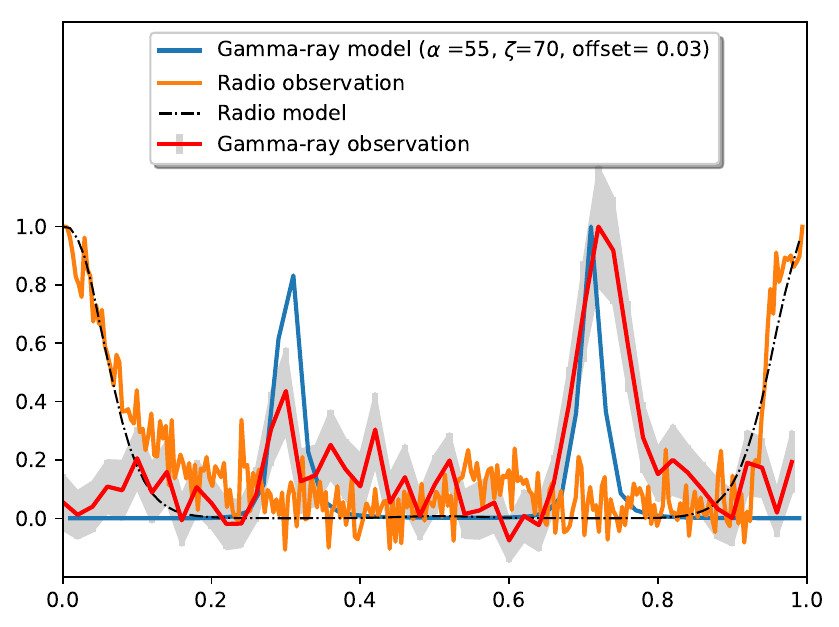}
    \caption{PSR~J0102+4839}
    \label{fig:J0102_bestfit}
  \end{subfigure}
  \begin{subfigure}[b]{0.4\linewidth}
    \includegraphics[width=0.95\linewidth]{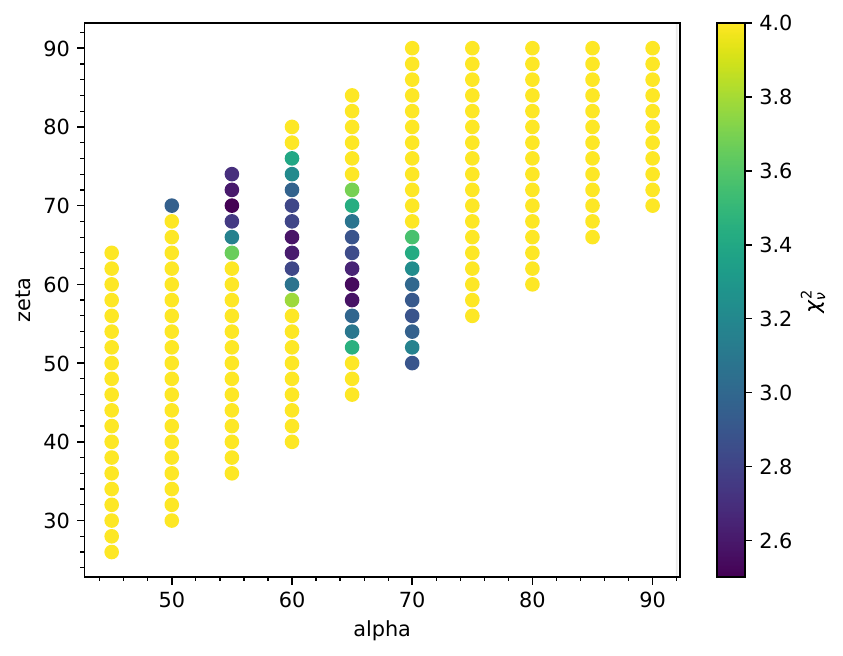}
    \caption{PSR~J0102+4839, $\phi = 0.03$ }
    \label{fig:J0102_chin}
  \end{subfigure}
  
  \begin{subfigure}[t]{0.4\linewidth}
    \includegraphics[width=0.95\linewidth]{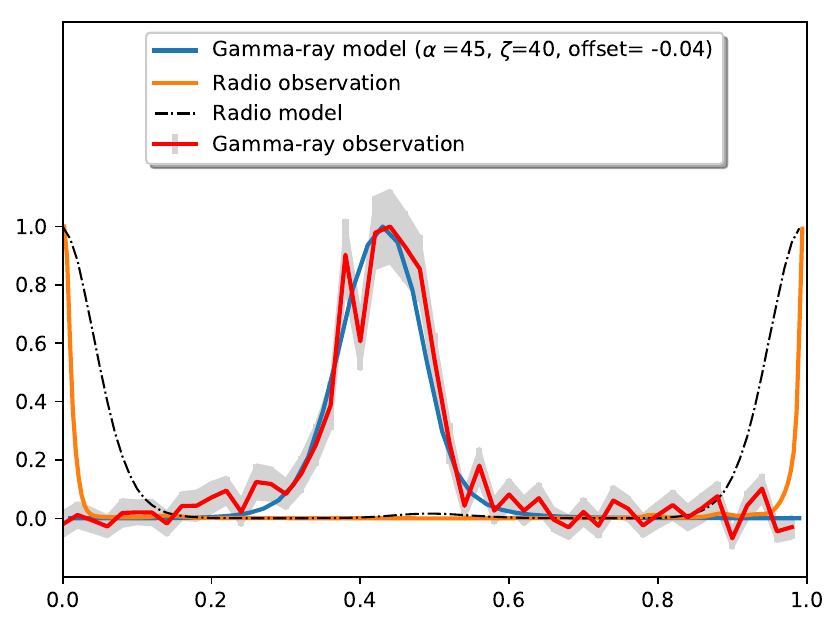}
    \caption{PSR~J0437--4715}
        \label{fig:J0437_bestfit}
  \end{subfigure}
  \begin{subfigure}[t]{0.4\linewidth}
    \includegraphics[width=0.95\linewidth]{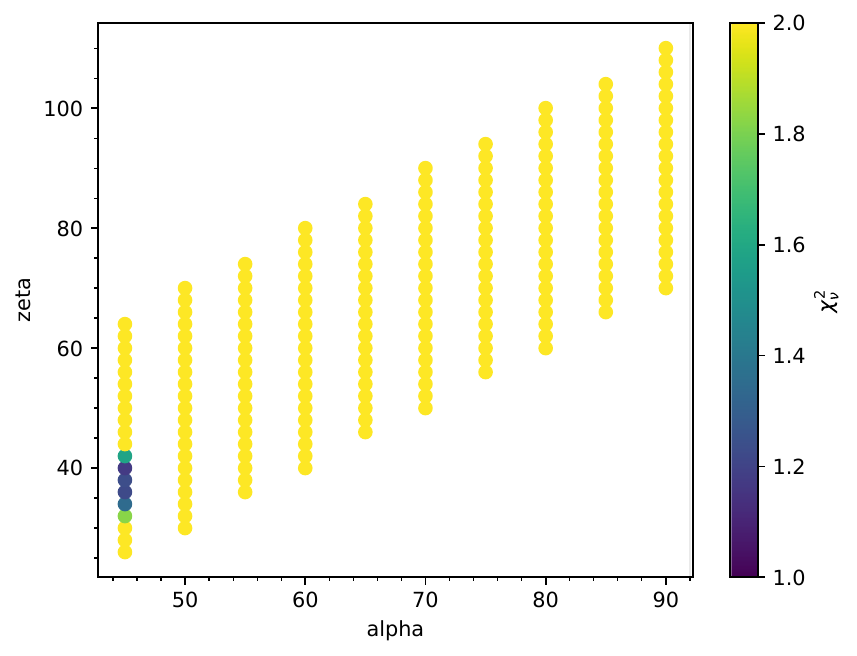}
    \caption{PSR~J0437--4715, $\phi = -0.04$ }
        \label{fig:J0437_chin}
  \end{subfigure}

    \begin{subfigure}[b]{0.4\linewidth}
    \includegraphics[width=0.95\linewidth]{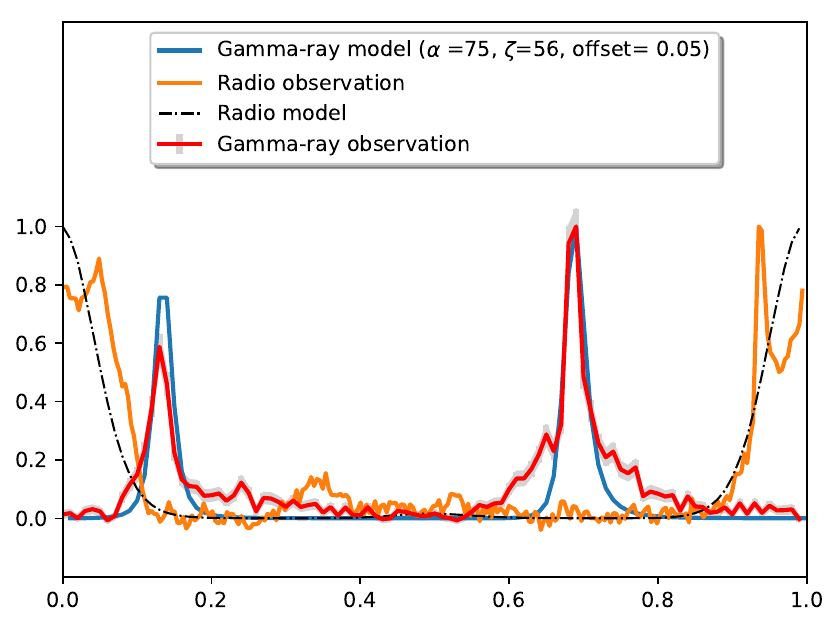}
    \caption{PSR~J0614--3329}
    \label{fig:J0614_bestfit}
  \end{subfigure}
      \begin{subfigure}[b]{0.4\linewidth}
    \includegraphics[width=0.95\linewidth]{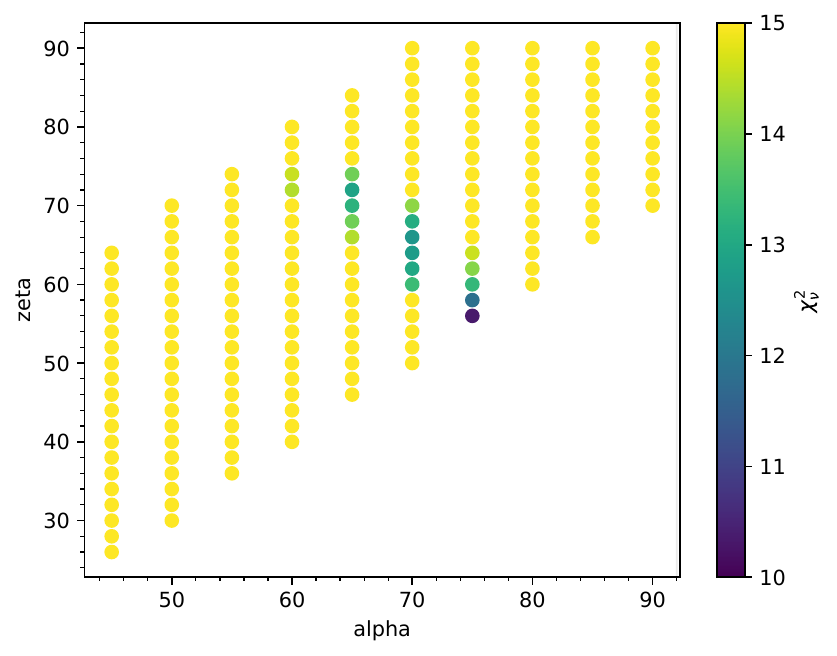}
    \caption{PSR~J0614--3329, $\phi = 0.05$ }
    \label{fig:J0614_chin}
  \end{subfigure}
  
  \caption{Best-fitting gamma-ray light curves (left panel) and reduced chi-square distributions (right panel).}
  \label{fig:bestfits_1}
\end{figure*}


\begin{figure*}[]
  \centering

    \begin{subfigure}[b]{0.4\linewidth}
    \includegraphics[width=0.95\linewidth]{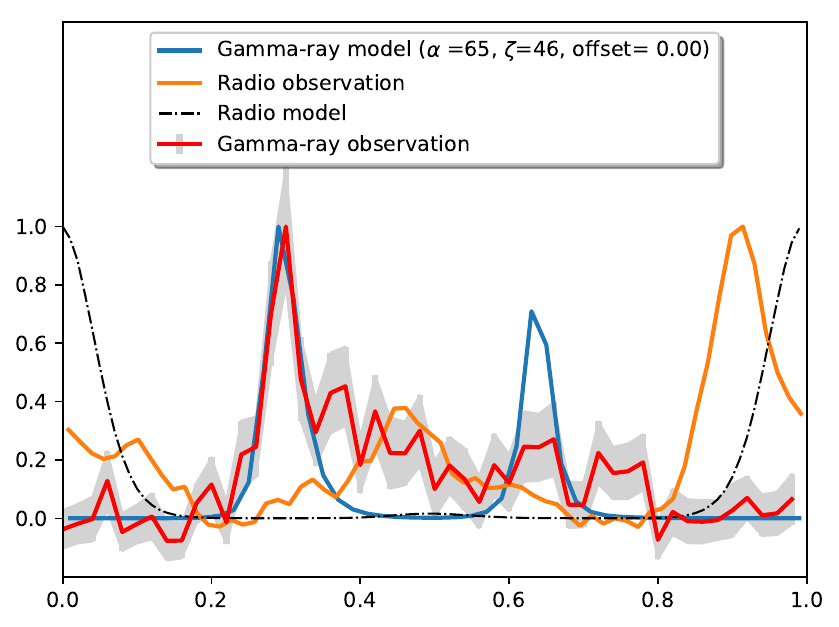}
    \caption{PSR~J1124--3653}
    \label{fig:J1124_bestfit}
  \end{subfigure}
    \begin{subfigure}[b]{0.4\linewidth}
    \includegraphics[width=0.95\linewidth]{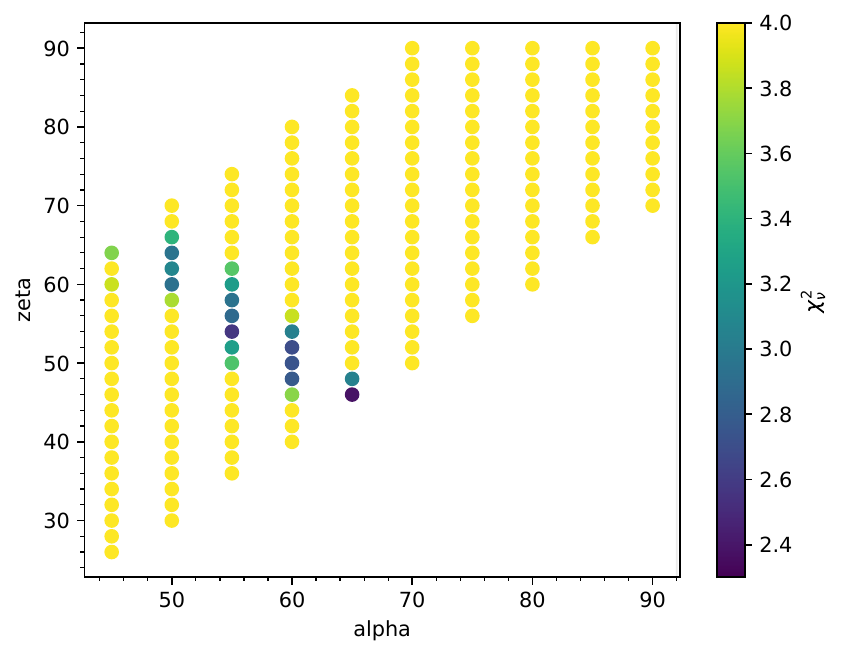}
    \caption{PSR~J1124--3653, $\phi = 0.00$ }
    \label{fig:J1124_chin}
  \end{subfigure}

    \begin{subfigure}[b]{0.4\linewidth}
    \includegraphics[width=0.95\linewidth]{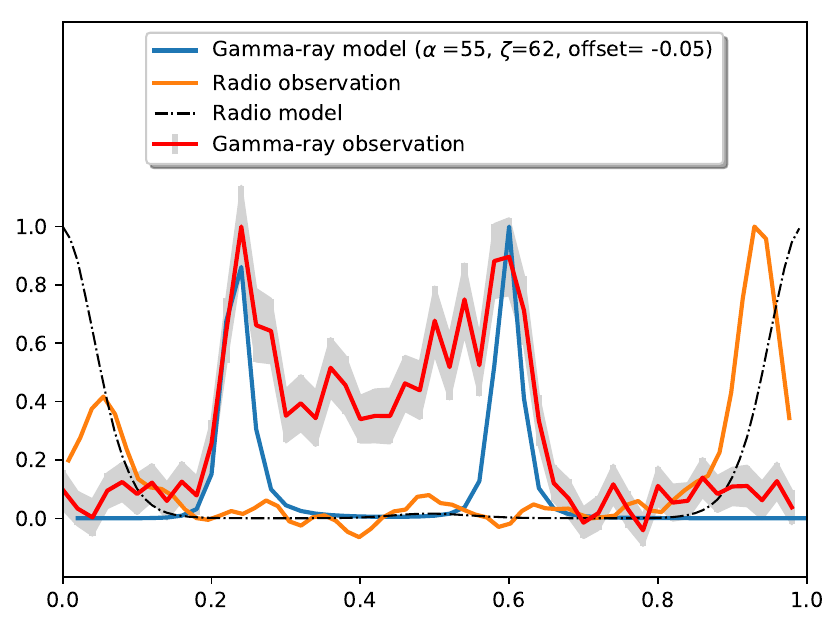}
    \caption{PSR~J1514--4946}
    \label{fig:J1514_bestfit}
  \end{subfigure}
      \begin{subfigure}[b]{0.4\linewidth}
    \includegraphics[width=0.95\linewidth]{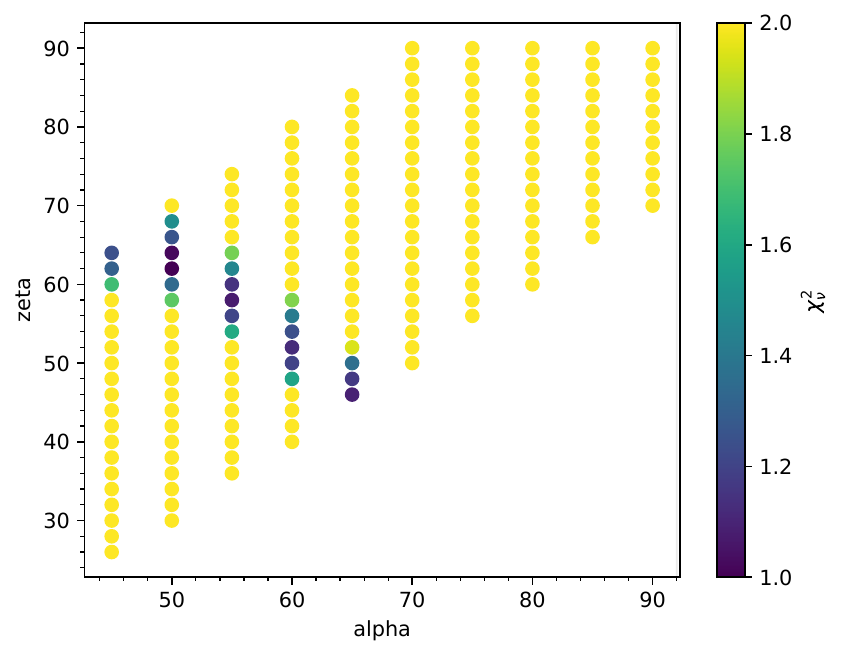}
    \caption{PSR~J1514--4946, $\phi = -0.05$ }
    \label{fig:J1514_chin}
  \end{subfigure}

  \begin{subfigure}[b]{0.4\linewidth}
    \includegraphics[width=0.95\linewidth]{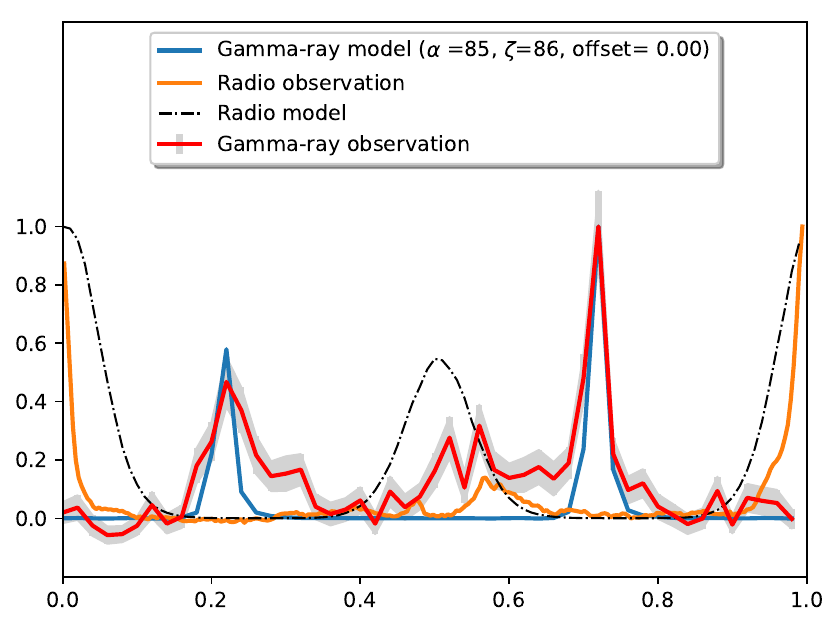}
    \caption{PSR~J1614--2230}
    \label{fig:J1614_bestfit}
  \end{subfigure}
  \begin{subfigure}[b]{0.4\linewidth}
    \includegraphics[width=0.95\linewidth]{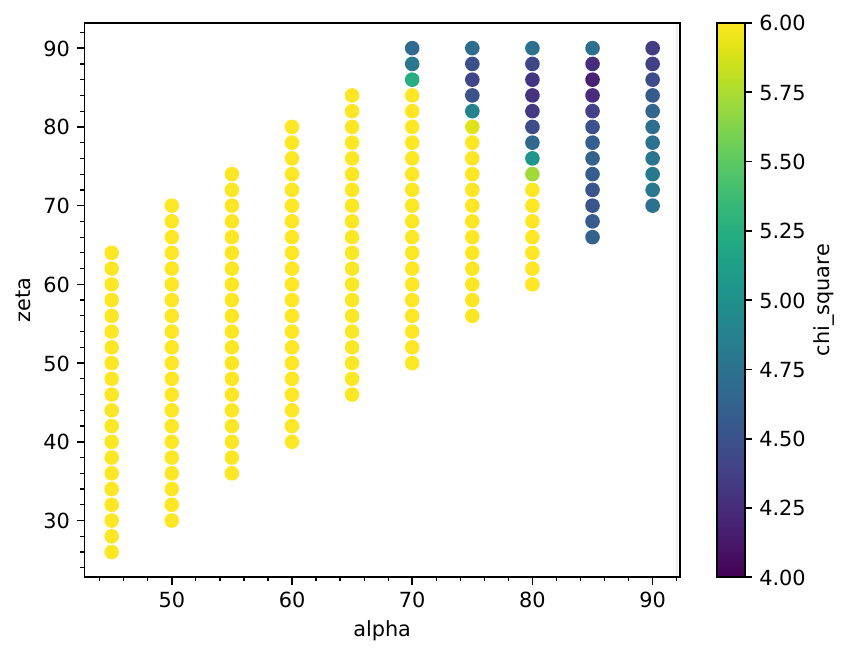}
    \caption{PSR~J1614--2230, $\phi = 0.00$ }
    \label{fig:J1614_chin}
  \end{subfigure}
  
  \begin{subfigure}[b]{0.4\linewidth}
    \includegraphics[width=0.95\linewidth]{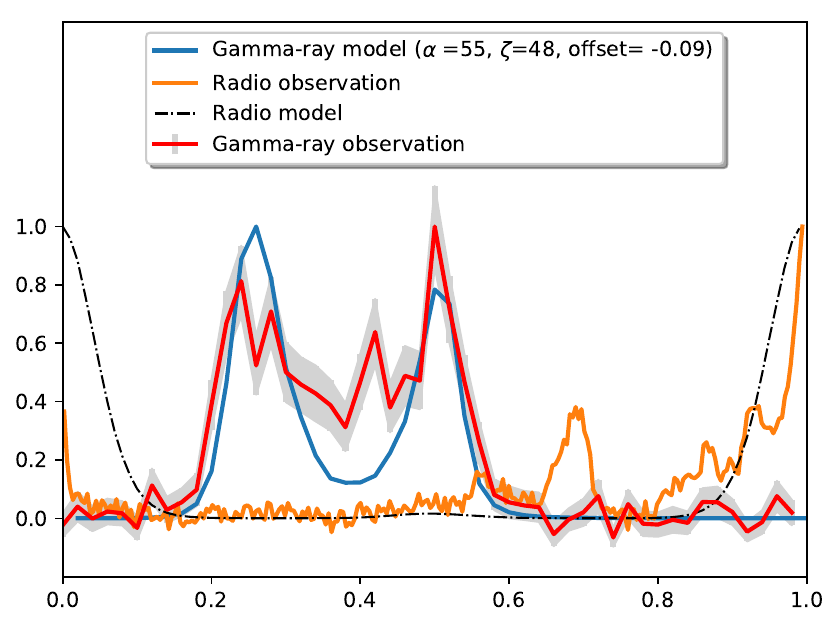}
    \caption{PSR~J2017+0603}
    \label{fig:J2017_bestfit}
  \end{subfigure}
  \begin{subfigure}[b]{0.4\linewidth}
    \includegraphics[width=0.95\linewidth]{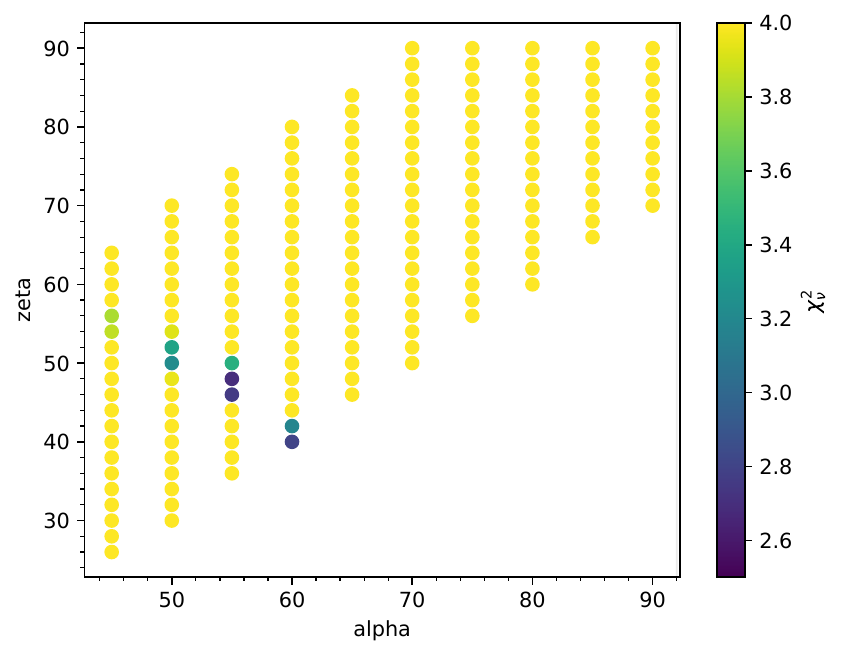}
    \caption{PSR~J2017+0603, $\phi = -0.09$ }
    \label{fig:J2017_chin}
  \end{subfigure}
  
  \caption{Best-fitting gamma-ray light curves (left panel) and reduced chi-square distributions (right panel).}
  \label{fig:bestfits_2}
\end{figure*}


\begin{figure*}[]
  \centering

  \begin{subfigure}[b]{0.4\linewidth}
    \includegraphics[width=0.95\linewidth]{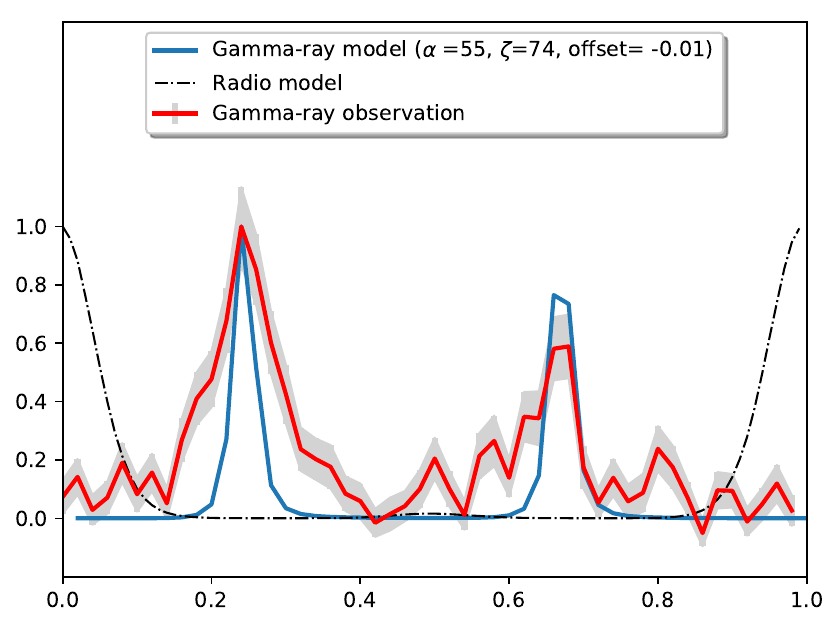}
    \caption{PSR~J2043+1711}
    \label{fig:J2043_bestfit}
  \end{subfigure}
  \begin{subfigure}[b]{0.4\linewidth}
    \includegraphics[width=0.95\linewidth]{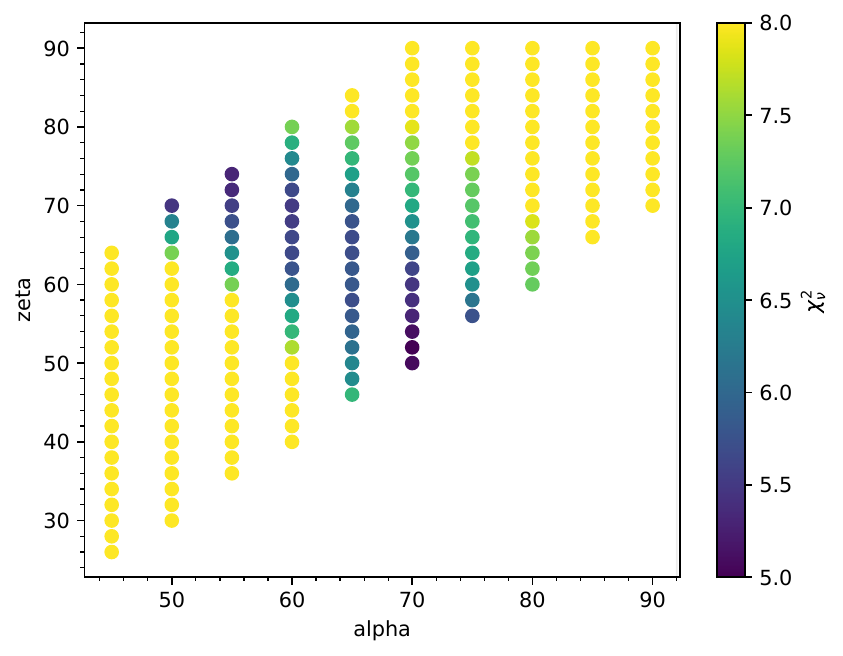}
    \caption{PSR~J2043+1711, $\phi = -0.01$ }
    \label{fig:J2043_chin}
  \end{subfigure}

  \begin{subfigure}[t]{0.4\linewidth}
    \includegraphics[width=0.95\linewidth]{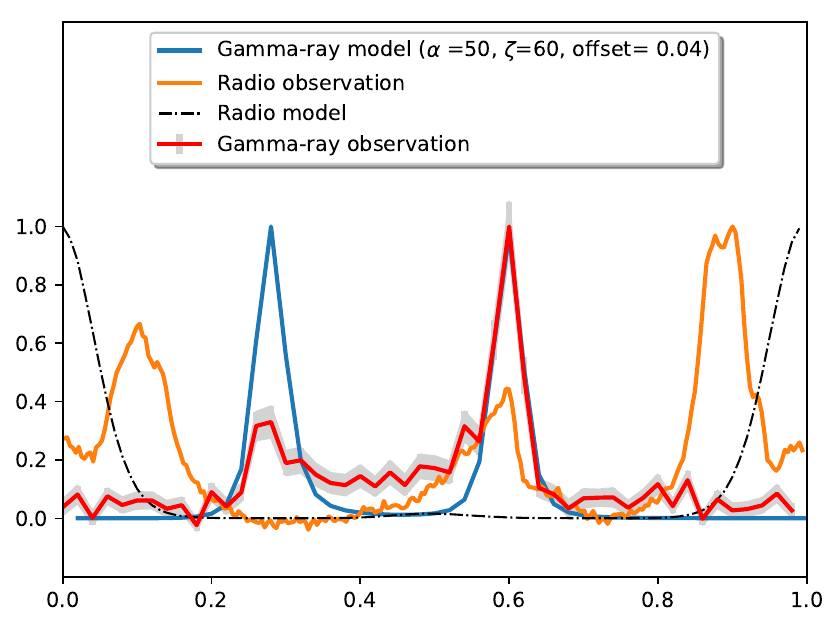}
    \caption{PSR~J2302+4442}
    \label{fig:J2302_bestfit}
  \end{subfigure}
  \begin{subfigure}[t]{0.4\linewidth}
    \includegraphics[width=0.95\linewidth]{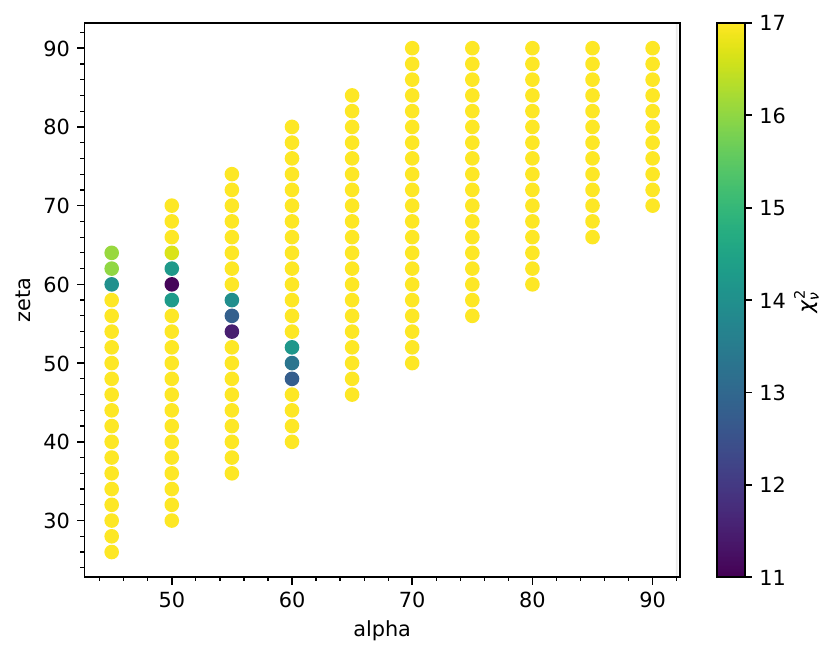}
    \caption{PSR~J2302+4442, $\phi = -0.03$ }
    \label{fig:J2302_chin}
  \end{subfigure}

  \caption{Best fitting gamma-ray light curves (left panel) and reduced chi-square distributions (right panel).}
  \label{fig:bestfits_3}
\end{figure*}


\begin{table*}
\caption{Estimated ranges of reasonable angles, $\alpha$, $\zeta$, by fits to the gamma-ray profiles with the artificial offsets, $\phi$, for the best fits. $\chin$ is the reduced chi-squared values for the best fits.}
\centering
\begin{threeparttable} 
\begin{tabular}{llllllll}
\hline
Source  Name                    & Period (ms)   &   $\alpha (^{\circ})$ & $\zeta (^{\circ})$ & $\alpha_\mathrm{Best Fit} (^{\circ})$ & $\zeta_\mathrm{Best Fit} (^{\circ})$ & $\phi$ (Period) & $\chin$    \\
\hline
PSR~J0030+0451                  & 4.87                  & 55 -- 75                              &       52 -- 74           & 70 & 60       &       -0.07                                           &       7.2     \\
PSR~J0102+4839                  & 2.96                  & 50 -- 70                              &       50 -- 76           & 55 & 70       &       0.03                                    &       2.5     \\
PSR~J0437--4715                 & 5.76                  & $\sim 45$                             &       34 -- 42           & 45 & 40       &       -0.04                                           &       1.2     \\
PSR~J0614--3329                 & 3.15                  & 65 -- 75                              &       56 -- 74           & 75 & 56       &       0.05                                    &       10.4 \\
PSR~J1124--3653                 & 2.41                  & 50 -- 65                              &       46 -- 66           & 65 & 46       &       0.00                                    &       2.4     \\
PSR~J1514--4946                 & 3.59                  & 45 -- 65                              &       46 -- 66           & 55 & 62       &       -0.05                                           &       1.0 \tnote{a} \\
PSR~J1614--2230                 & 3.15                  & 70 -- 90                              &       66 -- 90           & 85 & 86       &       0.00                                    &       4.2     \\
PSR~J2017+0603                  & 2.90                  & 45 -- 55                              &       40 -- 52           & 55 & 48       &       -0.09                                   &       2.7     \\
PSR~J2043+1711                  & 2.38                  & 50 -- 75                              &       46 -- 80           & 55 & 74       &       -0.01                                           &       5.3     \\
PSR~J2302+4442                  & 5.19                  & 45 -- 60                              &       48 -- 62           & 50 & 60       &       -0.03                                           &       11.1    \\
\hline
\end{tabular}
\begin{tablenotes}
\item[a] For PSR~J1514--4946  the best fits were retrieved by accounting for gamma-ray intensities at phases 0.2 -- 0.3 and 0.55 -- 0.65, not the entire phase.
\end{tablenotes}
\end{threeparttable}
\label{tab:parameters}
\end{table*}


\subsection{PSR~J0030+0451}

PSR~J0030+0451 is a MSP with $P = 4.87$~ms which shows double pulses in gamma-rays and has a relatively clear light curve profile in gamma-ray energy bands. The primary radio peak of the source shows a multi-peaked structure rather than a smooth single peak. The interpulse observed with much smaller amplitude compared to the main pulse implies a geometry close to an orthogonal rotator while the proximity of $\alpha$ to $90^{\circ}$ depends on the details of the radio emission procedure. As can be seen in Fig. \ref{fig:J0030_bestfit}, the best fit to the gamma-ray light curve of PSR~J0030+0451 was obtained by fixing $\alpha = 70^{\circ}$, $\zeta = 60^{\circ}$ and $\phi = -0.07$ giving a reduced chi-squared of $\chin = 7.2$. We find that $\alpha = (55^{\circ}-75^{\circ})$ and $\zeta = (52^{\circ}-74^{\circ})$ gives fits with $7 \lesssim \chin \lesssim 12$ (see Fig. \ref{fig:J0030_chin}).

\subsection{PSR~J0102+4839}

PSR~J0102+4839 is another MSP with $P = 2.96$~ms which has a primary peak and a second peak with lower amplitude in gamma-rays. Its radio profile has a broad single peak featured with broader trailing edge compared to the leading edge of the peak. We obtained the best fit to the gamma-ray profile with $\alpha = 55^{\circ}$, $\zeta = 70^{\circ}$, $\phi = 0.03$ and $\chin \simeq 2.5$; see Fig. \ref{fig:J0102_bestfit}. By taking $\phi = 0.11$, we calculated $\chin$ for different $\alpha$ and $\zeta$ values which indicates that reasonable fits can be obtained with $2.5 \lesssim \chin \lesssim 4$ for $\alpha = (50^{\circ}-70^{\circ})$ and $\zeta = (50^{\circ}-76^{\circ})$ (see Fig. \ref{fig:J0030_chin}).

\subsection{PSR~J0437--4715}

We analysed PSR~J0437--4715, the slowest MSP in our sample with $P = 5.76$~ms, which is an example of the sources showing radio pulsation and single peak in gamma-rays. Our results imply that the inclination angle is well constrained: $\alpha = 45^{\circ}$. The best fit is obtained by $\alpha = 45^{\circ}$, $\zeta = 40^{\circ}$, $\phi = -0.04$ with $\chin = 1.2$, see Fig.~\ref{fig:J0437_bestfit}, while $\alpha = 45^{\circ}$ and $\zeta = (34^{\circ}-42^{\circ})$ yield good fits within the range $1.2 \lesssim \chin \lesssim 2$ (Fig. \ref{fig:J0437_chin}). 

\subsection{PSR~J0614--3329}

PSR~J0614--3329, with $P = 3.15$~ms, has clear double peaks in the gamma-ray light curve. The amplitude of the leading peak is slightly lower than the trailing peak. The best fit to the gamma-rays with $\chin = 10.4$ is obtained with the parameters $\alpha = 75^{\circ}$, $\zeta = 56^{\circ}$ and $\phi = 0.05$; see Fig. \ref{fig:J0614_bestfit}. The fits with $10 \lesssim \chin \lesssim 15$ implies $\alpha = (70^{\circ} - 75^{\circ})$ and $\zeta = (60^{\circ}-70^{\circ})$ (see Fig. \ref{fig:J0614_chin}). In spite of high $\chin$ which is mostly due to the off-pulse emission, there is a strong match between the peaks produced by the model and the observations which is what we are looking for as our main goal.

\subsection{PSR~J1124--3653}

The gamma-ray profile of PSR~J1124--3653 ($P = 2.41$~ms) can be interpreted as either (1) having a single peak feature with a trailing edge gradually decreasing in intensity or (2)  having a less prominent second peak through the end of the trailing edge. We obtained the best fit to the gamma-ray profile with $\alpha = 65^{\circ}$, $\zeta = 46^{\circ}$ and $\chin = 2.4$ (see Fig. \ref{fig:J1124_bestfit}) while reasonable fits can be obtained with $2.4 \lesssim \chin \lesssim 4$ for $\alpha = (50^{\circ}-65^{\circ})$ and $\zeta = (46^{\circ}-66^{\circ})$; see Fig.~\ref{fig:J1124_chin}. We do not need to shift the model curve along the phase to be able to find a plausible fit.
We note that the amplitude of the second peak is overestimated in the best fit. 

\subsection{PSR~J1514--4946}

PSR~J1514--4946 ($P = 3.59$~ms) shows double peaks in gamma-rays with bridge emission in between these peaks. Our model does not address this inter-pulse emission. Trying to find best fits for the entire light curve at any phase did not work for this pulsar because of bridge emission. We therefore focused on the two main peaks. The best fit in Fig. \ref{fig:J1514_bestfit} was retrieved by accounting for the gamma-ray intensities at phases $0.2-0.3$ and $0.55-0.65$ which yielded a best fit with $\alpha = 55^{\circ}$, $\zeta = 62^{\circ}$, $\chin \simeq 1$ and $\phi = -0.05$ (Fig. \ref{fig:J1514_bestfit}). We found that reasonable light curves with $1 \lesssim \chin \lesssim 2$ can be retrieved for $\alpha = (45^{\circ} - 65^{\circ})$ and $\zeta = (46^{\circ}-66^{\circ})$ (Fig. \ref{fig:J1514_chin}).

\subsection{PSR~J1614--2230}

The best fit to the gamma-ray light curve of PSR~J1614--2230 ($P = 3.15$~ms) is shown in Fig. \ref{fig:J1614_bestfit}. It is obtained by choosing $\alpha = 85^{\circ}$, $\zeta = 86^{\circ}$ and $\phi = 0.00$ giving a $\chin = 4.2$. The reasonable fits with $7 \lesssim \chin \lesssim 12$ can be retrieved by the ranges $\alpha = (70^{\circ}-90^{\circ})$ and $\zeta = (66^{\circ}-90^{\circ})$ (Fig. \ref{fig:J1614_chin}). However, solutions with $\alpha \sim 90^{\circ}$ cannot represent the true geometry of the system because interpulse in radio was not observed for PSR~J1614--2230.

\subsection{PSR~J2017+0603}

PSR~J2017+0603 is a MSP with $P = 2.90$~ms which has a similar gamma-ray profile to that of PSR~J1514--4946, but with a narrower peak separation of $\Delta \sim 0.25$. The best fit is obtained by setting $\alpha = 55^{\circ}$, $\zeta = 48^{\circ}$ and $\phi = -0.09$ with $\chin \simeq 2.7$; see Fig. \ref{fig:J2017_bestfit}. The geometry could be relatively well constrained for this pulsar with $\alpha = (45^{\circ} - 55^{\circ})$ and $\zeta = (40^{\circ}-52^{\circ})$ by the presumption that the model curves with $2.7 \lesssim \chin \lesssim 4$ are good candidates to represent the observed light curve (Fig. \ref{fig:J2017_chin}).

\subsection{PSR~J2043+1711}

Of the sources examined in this study, PSR~J2043+1711 is the fastest spinning pulsar with $P = 2.38$~ms. The amplitudes and locations of both pulsations in the gamma-ray are well fitted with our model. We obtained the best fit by setting $\alpha = 55^{\circ}$, $\zeta = 74^{\circ}$ and $\phi = -0.01$ with $\chin = 5.3$ while the reasonable model curves could be found by the angles in the range $\alpha = (50^{\circ} - 75^{\circ})$ and $\zeta = (46^{\circ}-80^{\circ})$ for $5.3 \lesssim \chin \lesssim 8$; see Fig. \ref{fig:J2043_chin}. We want to point out that we do not show the observed radio light curve of the source because we have not found any radio data in the 2PC files corresponding to this pulsar. However, it is not directly related to our analysis here as we only deal with the time-alignment of gamma-ray and radio not the exact shape of the radio profile.

\subsection{PSR~J2302+4442}

PSR~J2302+4442, with a period of $P = 5.19$~ms, shows double peaks in the gamma-rays. The best fit is produced with $\alpha = 50^{\circ}$, $\zeta = 60^{\circ}$ and an offset $\phi = -0.03$ with $\chin = 11.1$; see Fig.~\ref{fig:J2302_bestfit}. The ranges of the angles, provided that $11.1 \lesssim \chin \lesssim 17$, are found to be $\alpha = (45^{\circ} - 60^{\circ})$ and $\zeta = (48^{\circ}-62^{\circ})$; see Fig.~\ref{fig:J2302_chin}. In the best fit curve, the second peak perfectly matches the profile of the observed second peak. Although we overestimate the amplitude of the first peak, the peak locations are in line with the observations.

\section{Discussion}
\label{sec:Discussion}

In this study, we focused on the relationship between the radio time lag, the gamma-ray peak separation, the ratio of the peak amplitude to obtain the geometric parameters of pulsar magnetospheres, the obliquity, and the inclination of the line of sight. 
However, there is some uncertainty as to the precise location of the radio and high-energy emission sites. This leads to possible variations in the time lag between radio and gamma-rays. For instance, retardation effects within the light cylinder due to the finite propagation speed of light amount to an additional phase shift of \begin{equation}\label{eq:retarded_phase}
\phi_{\rm ret} = \frac{\Delta r}{2\,\pi\,\rlight} \lesssim \frac{1}{2\,\pi} \approx 0.16 .
\end{equation}
This shift has been implemented in our approach by adding a phase $\phi$ that also corrects for the period of MSP being longer than the 1.2~ms used for the force-free simulations and the emission sky maps. An additional unknown comes from the observation side. Indeed, there is no hint for the maximum of the radio peak to represent the centre of the pulse profile. It could well be that one or several cone emission patterns produce the pulses with sub-pulses of different intensities. In our simulations, we exactly know the location of the magnetic axis and the associated polar cap centre. Therefore, we used the radio peak as phase zero for our simulations. Nevertheless, for the observational data, there is no hint of a one-to-one correspondence between peak intensity and the middle of the pulse profile. This effect adds another unconstrained phase-shift between radio and gamma-rays. Further, while estimating the radio time lag~$\delta$, we assumed the radio emission to originate from regions of open dipolar magnetic field lines. However, estimates of $\delta$ will be affected if the radio emission arises from regions where there is an influence of the strong non-dipolar magnetic field.

From a more fundamental physical perspective, the microphysics of pair creation, acceleration, and their outflow into the striped wind is still  poorly understood. The particle density number, its energy distribution function, and their radiation spectra are not known with any accuracy. Nevertheless, the wind Lorentz factor, the current sheet thickness, and the flow velocity pattern all affect the gamma-ray pulse profiles. It is difficult to estimate the emissivity of the wind without the full description of the particle and radiation dynamics from the stellar surface to the wind. Clearly, further physical insight is required in order to precisely fit the pulse shape for individual pulsars. Such attempts have been avoided in this study because of the number of uncertainties in the model. Our approach, although simple, deals with the least number of parameters but already satisfactorily fits existing light curves. There is no doubt that more extensive works, including a kinetic description of the magnetosphere and its associated radiation mechanisms, will unveil many important fundamental issues in pulsar physics.

\section{Conclusion}
\label{sec:Conclusion}

We numerically solved the equations for force-free pulsar magnetospheres for many obliquities~$\alpha$ and computed the corresponding gamma-ray pulsed emission emanating from the striped wind. Through individual analyses of ten MSPs with spin periods in the range~$P \sim (2 - 6)$~ms, we showed that their gamma-ray pulse profiles could be faithfully reproduced by assuming that radio pulses escape from open field-line regions close to the polar caps. In contrast, gamma-ray pulses are produced in the current sheet of the striped wind, outside the light cylinder. In this study, we constrained the range of parameters describing the geometry of each pulsar, namely their magnetic and light of sight inclination angles with respect to the spin axis of the star.

In order to retrieve good fits to the observed gamma-ray light curves of each source, we applied the least-square method to compare the 
reported intensities with those found from the model in each bin of the observation throughout one complete rotation of the star. We present our estimations together with their statistics in Table~\ref{tab:parameters}. We were able to fit all pulsars correctly.

The up-to-date measurement of radio pulse profiles and PPA traverse of MSPs indicates extremely irregular behaviour. Therefore, we were not able to use such techniques to deduce independent knowledge of their geometry. The distortion of the PPA can result from altitude-dependent aberration retardation effects \citep{mitra_effect_2004} or the presence of non-dipolar surface magnetic fields, which renders the PPA analysis useless for constraining $\alpha$ and $\zeta$.

We plan to apply the same technique to young pulsars which possess much more well-defined radio pulse profiles and PPA. Indeed, thanks to the RVM model, the radio emission height has been located at about 5\% of the light-cylinder radius \citep{mitra_nature_2017}. Even though such heights are much smaller than $\rlight$, they are large enough compared to the stellar radius, thus the dipole field is dominant. The force-free dipolar magnetosphere is therefore a good approximation for the computation of the observational signature in radio and gamma-rays. There is no more freedom for the radio altitude, and because of the narrow radio peaks, the uncertainties on the radio zero phase are no longer critical for fitting the $\delta-\Delta$ relation with high confidence. However, as young pulsars do not fall into the same category as MSPs, we will show the results in a separate forthcoming work.

The most recent and best-quality published gamma-ray data of pulsars were reported in the second Fermi Catalogue in 2013. Since then, almost one decade has passed and more observations have accumulated, increasing the signal-to-noise ratio of gamma-ray light curves and with improved intensity resolution. These updates are expected to be published as the third Fermi Pulsar Catalogue (3PC) in the future, which will undoubtedly increase the number of gamma-ray pulsars and provide even more stringent constraints on gamma-ray data to be tested with our model.

\bibliographystyle{aa}
\bibliography{Ma_bibliotheque.bib} 

\section*{Acknowledgements}

We are grateful to the referee for helpful comments and suggestions. This work is supported by the CEFIPRA grant IFC/F5904-B/2018. We would like to acknowledge the High-Performance Computing Centre of the University of Strasbourg for supporting this work by providing scientific support and access to computing resources. Part of the computing resources was funded by the Equipex Equip@Meso project (Programme Investissements d'Avenir) and the CPER Alsacalcul/Big Data. DM acknowledge the support of the Department of Atomic Energy, Government of India, under project no. 12-R\&D-TFR-5.02-0700.

\end{document}